\documentclass[%
reprint,
amsmath,amssymb,
aps,
]{revtex4-2}
\usepackage{graphicx}
\usepackage{dcolumn}
\usepackage{bm}
\usepackage{amsmath}
\usepackage{physics}
\usepackage{hyperref}
\usepackage{float}
\begin{document}
\preprint{APS/123-QED}
\title{Belief System Dynamics as Network of Single Layered Neural Network}
\author{Yujian Fu}
\affiliation{%
Department of Physics, University of Michigan, Ann Arbor, Michigan 48109, USA
}%
\affiliation{%
Center for the Study of Complex Systems, University of Michigan, Ann Arbor, Michigan 48109, USA}%
\begin{abstract}
    As problems in political polarization and the spread of misinformation become serious, belief propagation on a social network becomes an important question to explore. Previous breakthroughs have been made in algorithmic approaches to understanding how group consensus or polarization can occur in a population. This paper proposed a modified model of the Friedkin-Johnsen model that tries to explain the underlying stuborness of individual as well as possible back fire effect by treating each individual as a single layer neural network on a set of evidence for a particular statement with input being confidence level on each evidence, and belief of the statment is the output of this neural network. 

    In this papar, we reafirmed the importance of Madison's cure for the mischief of faction, and found that when structure of understanding is polarized, a network with a gaint component can decrease the variance in the belief distribution more than a network with two communities, but creates more social pressure by doing so. We also found that when community structure is formed, variance in the belief distribution become less sensitive to confidence level of individuals. The model can have various applications to political and historical problems caused by misinfomation and conflicting economic interest as well as applications to personality theory and behavior psychology. 
\end{abstract}
\maketitle

\section{Introduction}
Contributions have been made to understand the underlying mechanism of belief polarization and group consensus. Some examples in mathematical sociology include the introduction of bias assimilation by Dandekar et al. or stubborness by Friedkin and Johnsen into the DeGroot model\cite{Dandekar2013}\cite{Friedkin1990}\cite{Friedkin1999}. 

Inspired by Friedkin's model with logic constraint and Madalina's application of BENDING model in Psychology\cite{Friedkin2016}\cite{Vlasceanu2024}, a richer cognitive structure behind the formation and maintanence of an individual's belief is introduced. Statements of logically related beliefs are often interconnected and evidence plays a role in the formation of belief just as important as perceived norm. 

For the same idea, different people can have different ways of understanding it. Two people who believe in the same thing might update their belief differently after experiencing an event that updates their evidence in the same way. For example, consider the statement: I want to study physics. Some people might decide to do physics because they like solving challenging problems faster than others, some people might believe in it because they like discovery and beauty. Both types of people would succeed in a physics class, but when facing a long term research question only one out of the two can persevere through.

This paper proposed an alternative model of belief propagation where an individual's belief on a statement is decomposed into two components: self-reasoning, and social norm. Self-reasoning is a weighted sum over a pool of evidence that supports an individual to agree on a specific statement, and social norm is a weighted sum over the belief of an individual's neighbors. The weights on self-reasoning has the same interpretation as the stubborness value in the Friedkin-Johnsen model. For the purpose of this paper we are going to call the stubborness value in the Friedkin-Johnsen model, self-confidence, because it is the level of confidence an individual have on his or her own structure of understanding. By introducing evidence and structure of udnerstanding as an additional underlying structure of belief, we are able to reproduce the effect of logic constrain on multiple statements as their shared set of evidence and implementing BENDING model numerically.

This model also tries to explain backfire effect of belief propagation by introducing opposing evidence. For example, some people like chocolate because it is sweet, but some people like chocolate because it is bitter. But bitter and sweet are opposite ideas that supports the same statement --- chocolate is good. When an individual receives an evidence that is opposite to one of the evidence considered by the person's structure of understanding, instead of increasing the certainty this person has on the belief, it decreases the belief.

This model also measures the social pressure on an individual as the difference between their resulting belief and self-reasoning. At each timestep, confidence level on each evidence is propagated in the social network according to DeGroot model, and belief is calculated with a weighted sum over the evidence and neighbors' beliefs.

The example of chocolate preference is an example of how structure of understanding can be formed based on pure biological preference that is logical in the point of view of the individual who holds them, but may not in other people's point of view. There can be multiple different factors contributing to the formation of different structure of understanding including culture, socioeconomic status, level of education, experience in childhood... Everyone learns about the world in their unique way, and everyone can be ignorant of their own bias. Structure of understanding is not fixed. It can change over time based on experience or sometimes even randomly. In this paper, we are interested in belief formation based on the structure of understanding of people at a fixed time.

To make the theory general, this paper keeps the statment arbitrary with an arbitrary set of evidence. By choosing various statements, the theory can potentially have applications in political science, history, developmental psychology, personality theory and behavioral psychology by interpreting the structure of understanding as weights on different motivations for actions. 

Results of belief distribution and social pressure are found under assumptions of different population size, network geometry, self-confidence level and structure of understanding. 

\section{Methods}
Specific implementation of the model is described in the sections below. 

\subsection{Model Environment}
\begin{figure*}
    \centering
    \includegraphics[width=0.65\linewidth]{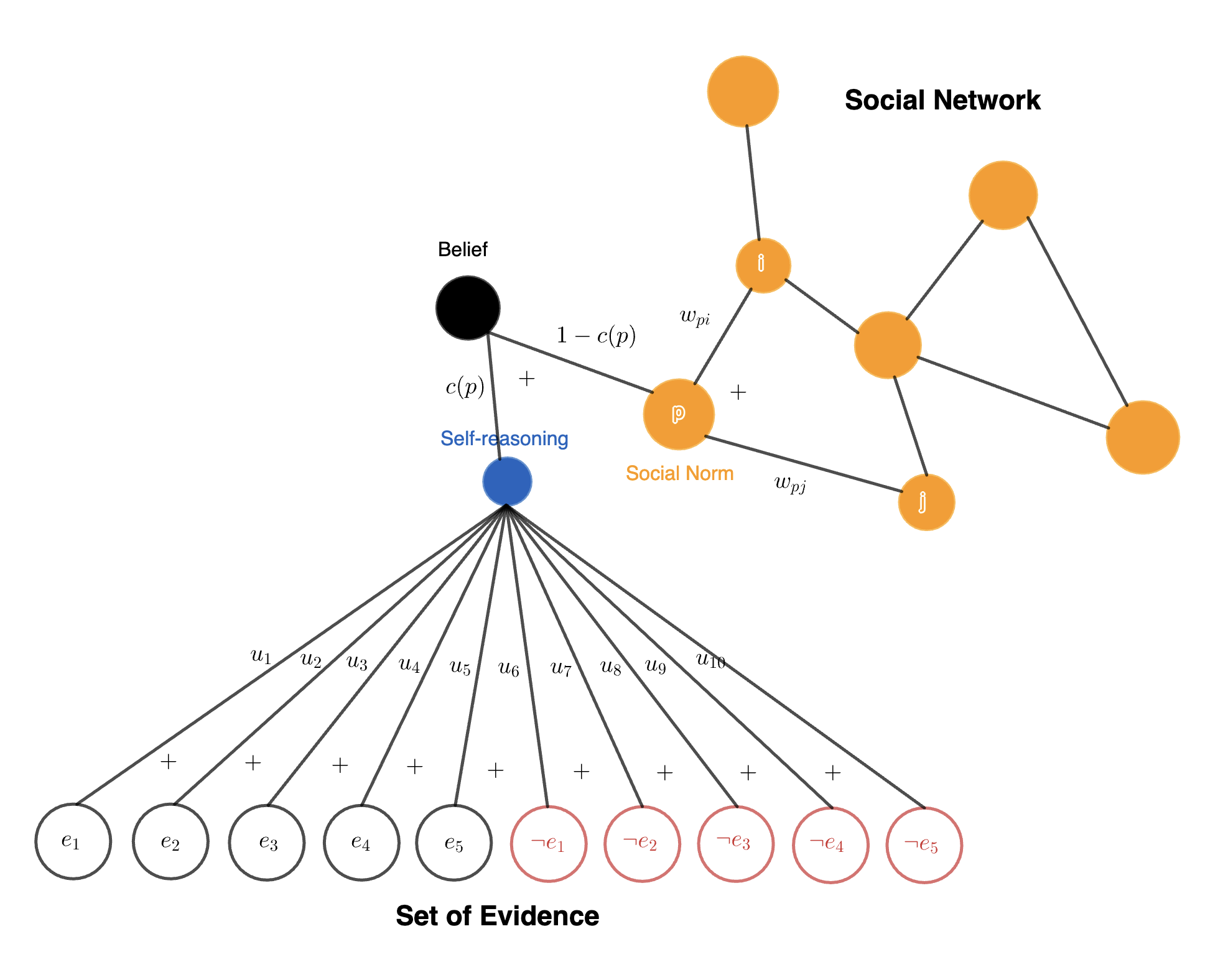}
    \caption{A diagram illustration of the model framework for the belief formation and propagation of agnet $p$. A belief is decomposed in to social norm, which is a weighted sum of perceived beliefs of the neighbors and self-reasoning, which is a weighted sum over a set of positive/negative evidence pair.}
    \label{fig:enter-label}
\end{figure*}
We initialize the model environment with two objects:
\begin{itemize}
    \item[1.] \textbf{Network}: The social network where the agents interact is defined to be a Erdős–Rényi random graph with probability of an edge between any two nodes equals to $\frac{k}{n}$, where $n$ is the size of the population and $k < n$ is the connectivity index, which describes the average number of people an individual is connected to. The adjacency matrix of the network, $\{w_{ij}\}$ is made doubly stochastic to ensure the sum of weights for all the edges of a node to be $1$. The network can be generated with two different options:
    \begin{itemize}
        \item[a.] One giant component
        \item[b.] Two communities
    \end{itemize}
    \item[2.] \textbf{Evidence pool}: The pool of evidence of length $2m$ for a specific statement is constructed by $m$ statements followed by $m$ negation of those statements. Structure of understanding of an individual would be constructed by $m$ nonzero weights choosing from the $m$ positive/negative evidence pairs and the other $m$ weights to be zero. Definition of the structure of understanding is given in the agent property section below. 
\end{itemize}
After initializing these two main objects, the model then read in pregenerated files that store information to initialize the agent properties to set up the simulation. 

\subsection{Agent Properties}
Over the course of a simulation, each agent stores these five information in their memory:
\begin{itemize}
    \item[1.] \textbf{Structure of understanding}: For an individual $p$, the structure of understanding of the individual on a particular statement is defined to be the set of weights \(0 \le u_i(p) \le 1\) corresponding to the evidence \(e_i\) on supporting the statement that satisfies $\sum_j^{2m} u_j(p) = 1$. Structure of understanding can be constructed in two different options:
    \begin{itemize}
        \item[a.] Randomly generated
        \item[b.] Lean towards positive side of the evidence pool with probability $p$ and towards negative side of the evidence pool with probability $1-p$. Where $p$ is defined to be the polarization index.
    \end{itemize}

    \item[2.] \textbf{Confidence level on evidence}: For an individual $p$, the set of confidence level on evidence \(0 \le b_i(p) \le 1\) is the level of certainty on a particular piece of evidence to be true. Where $1$ means maximal agree and $0$ means maximal disagree. When initializing the confidence level on evidence, one can also construct it in two different options:
    \begin{itemize}
        \item[a.] Randomly generated
        \item[b.] Positive evidence have initial confidence $a$ and negative evidence have initial confidence $1-a$.
    \end{itemize}

    \item[3.] \textbf{Self-confidence level}: For an individual $p$, the self-confidence level $0 \le c(p) \le 1$ is the weight contribution of self-reasoning to the belief. $1-c(p)$ is therefore the weight contribution of social norm to the belief.

    \item[4.] \textbf{Belief}: For an individual $p$, the belief $X(p)$ on a particular statement is calculated as follow:
    \begin{align*}
        X(p) &= c(p)\sum_{i} b_{i}(p)u_i(p)\\
        &+ (1-c(p)) \sum_{k\in N(p)} X(k)w_{kp}
    \end{align*}
    where $N(p)$ is the set of neighbors of person $p$. $w_{kp}$ is the weight of the edge between person $k$ and $p$. 
    \item[5.] \textbf{Social pressure}: For an individual $p$, the social pressure $P(p)$ on a particular statement is calculated as follow:
    \begin{align*}
        P(p) = \abs{X(p) - \sum_{i} b_{i}(p)u_i(p)}
    \end{align*}
\end{itemize}

\subsection{Evidence propagation}
At each step, a person will update their level of confidence on an evidence using a weighted sum over each of its neighbor's level of confidence. For example, at time \(t\), \(b_k(i)\) and \(b_l(i)\) are being communicated by person \(i\) to person \(p\). If \(u_k(p) = u_l(p) = 0\), then nothing happens to \(b_k(p), b_l(p)\). If at least one of them are not, WOLOG \(u_k(p) \ne 0\), then 
\[
b_k(p) = (1-w_{ip})b_k(p)+ w_{ip}b_k(i)
\]
If both of them are not, then 
\begin{align*}
    b_k(p) &= (1-w_{ip})b_k(p)+ w_{ip}b_k(i)\\
    b_l(p) &= (1-w_{ip})b_k(p)+ w_{ip}b_l(i)
\end{align*}
If \(e_k\) and \(e_l\) are \textbf{opposite evidence}, and $b_k(i)$ is being communicated by person $i$ to person $p$, then instead of adjusting the new level of certainty based on the level of certainty of the neighbor's belief, we adjust based on its complement. In addition to updating the confidence level $b_k(p)$, we also need to update $b_l(p)$ according to
\begin{align*}
    b_l(p) = (1-w_{ip})b_k(p)+ w_{ip}(1 - b_k(i))
\end{align*}

\section{Experiments}
\begin{figure*}
\begin{minipage}{0.45\textwidth}
        \centering
        \includegraphics[width=\textwidth]{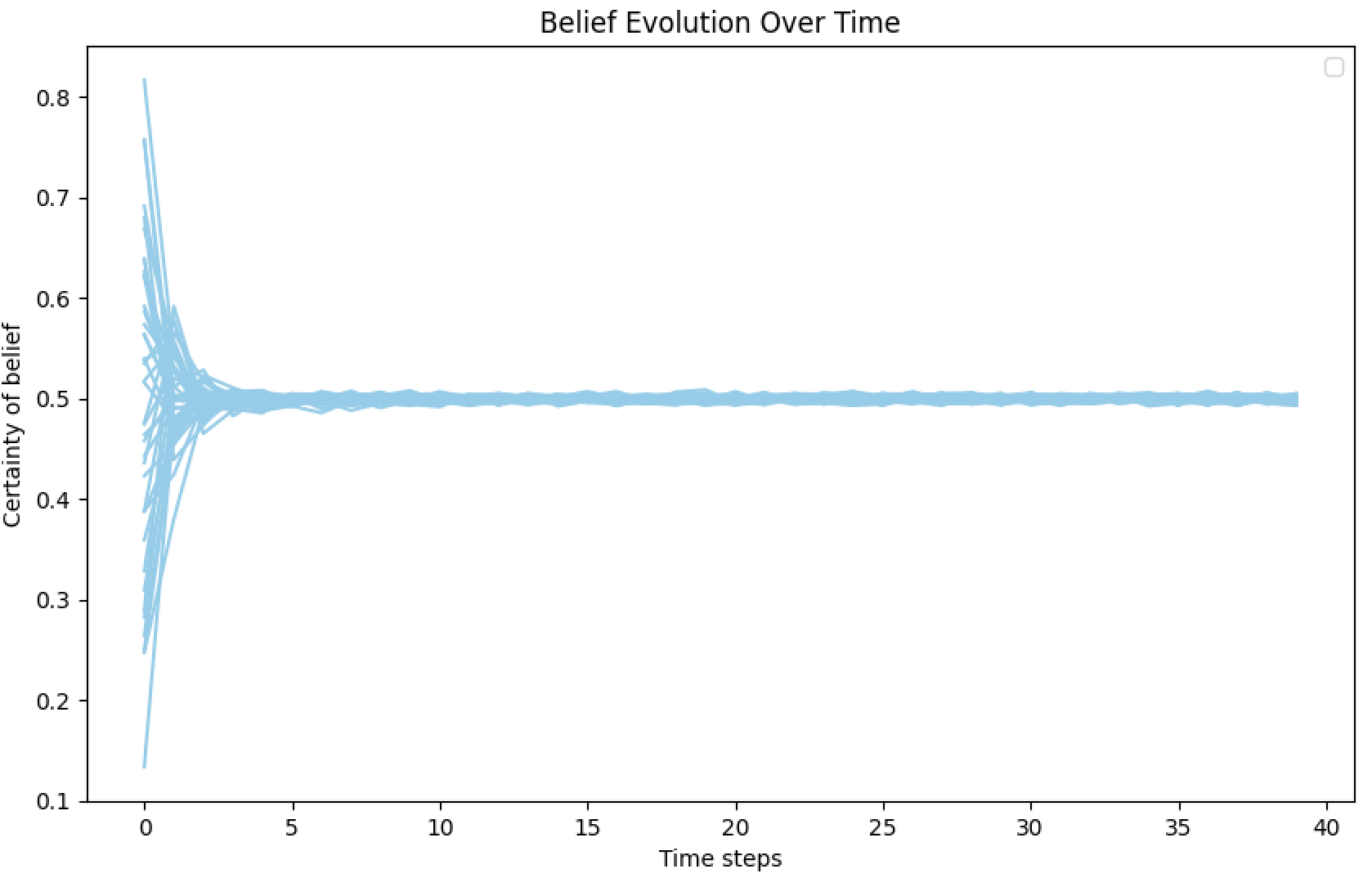} 
    \end{minipage}
    \begin{minipage}{0.45\textwidth}
        \centering
        \includegraphics[width=\textwidth]{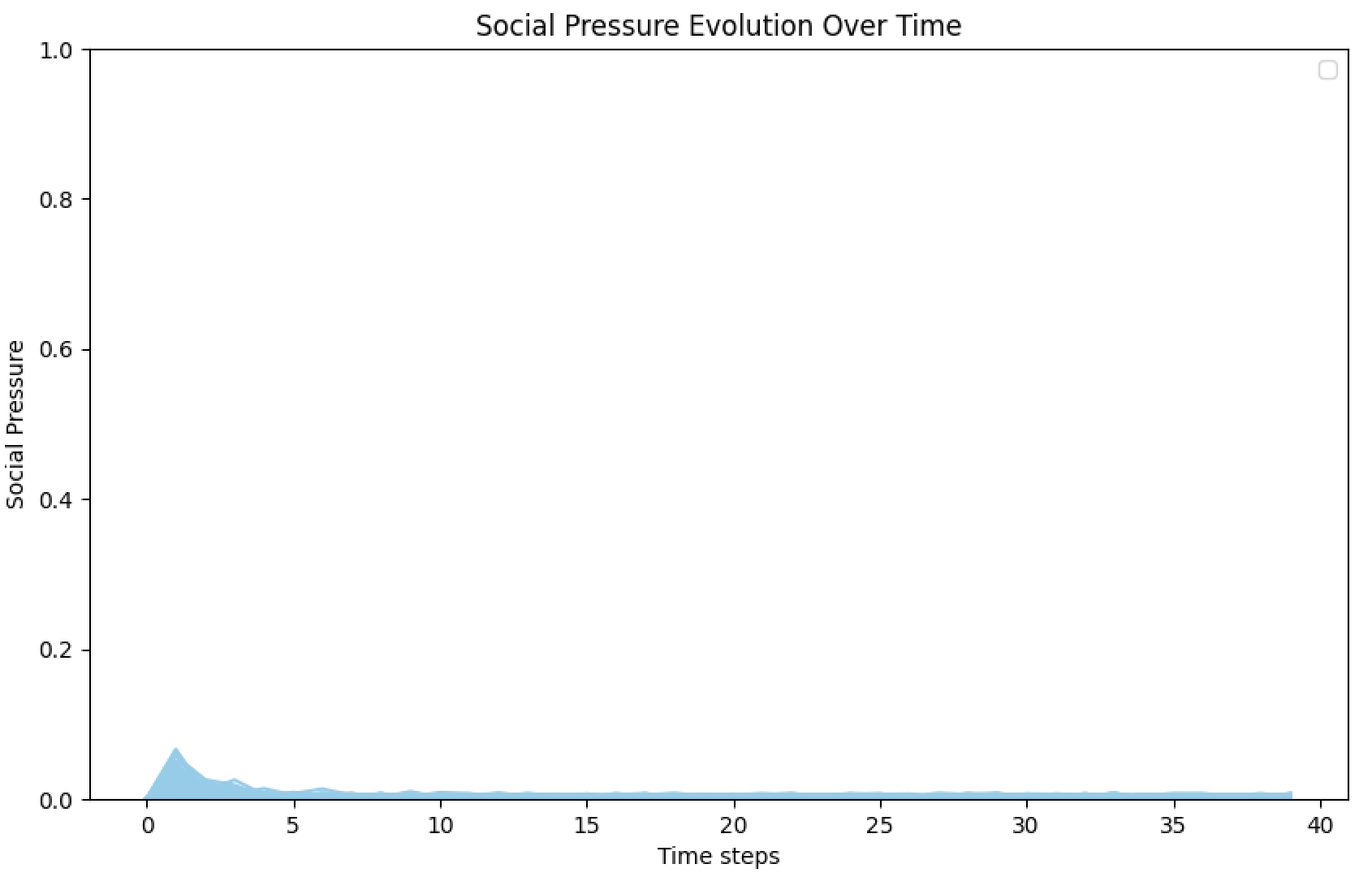} 
    \end{minipage}
    \caption{Evolution of beliefs (left) and social pressure (right) of the first 40 agents over time for randomly generated structure of understanding and initial confidence level.}
    \label{fig:enter-label}
\end{figure*}
\begin{figure*}
    \centering
    \begin{minipage}{0.3\textwidth}
        \centering
        \includegraphics[width=\textwidth]{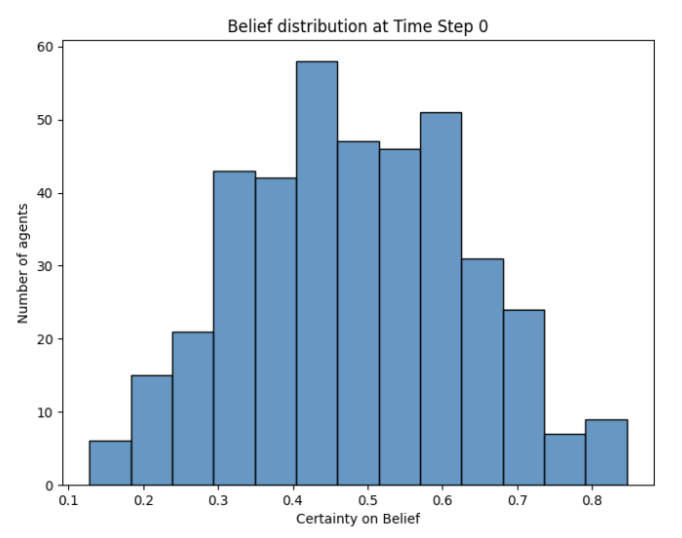} 
    \end{minipage}
    \begin{minipage}{0.3\textwidth}
        \centering
        \includegraphics[width=\textwidth]{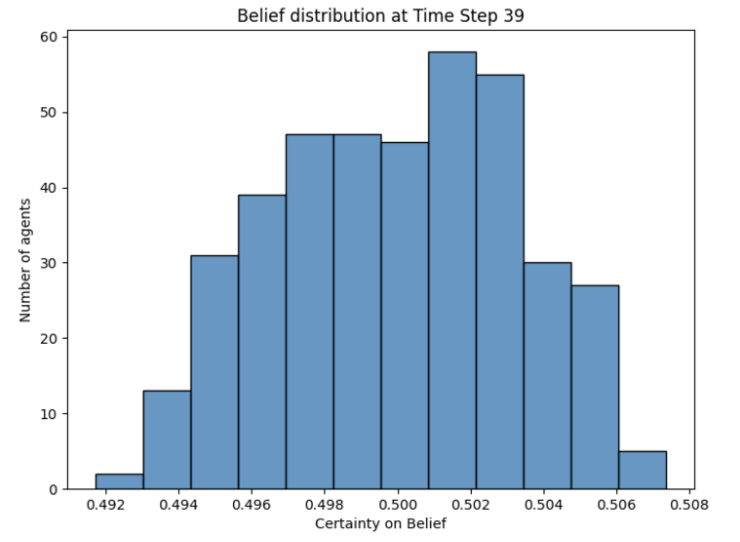} 
    \end{minipage}
    \begin{minipage}{0.3\textwidth}
        \centering
        \includegraphics[width=\textwidth]{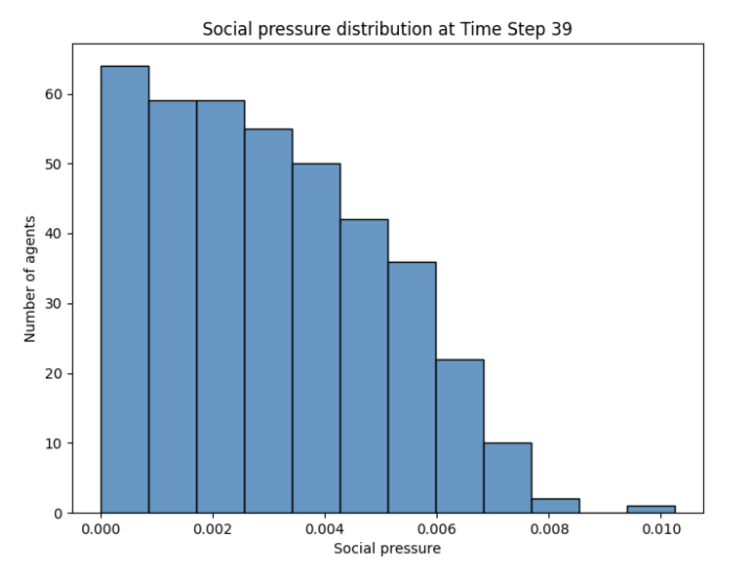} 
    \end{minipage}
    \label{fig:enter-label}
    \caption{Initial and final belief distribution and social pressure distribution}
\end{figure*}

The experiment consists of three trials, the first trial is on randomly initiated structure of understanding and initial confidence levels on a network with giant component; second trial is on polarized structure of understanding and initial confidence levels on a network with community structure; third trial is on polarized structure of understanding and initial confidence levels on a network with giant component. At each trial, network visualization, initial and final belief distribution will be plotted as well as social pressure. Investigation on the realtionship between standard deviation in the belief distribution with other variables is also shown.

The population size is usually set at $n = 400$, number of positive evidence is usually set at $m = 5$, self-confidence level is set at $0.5$, and connectivity is set at $k=10$ to allow majority of the nodes to be inside a giant component in the Erdős–Rényi random graph. For each simulation we run 40 time steps, which ensures convergence for cases considered. 

\subsection{Random Initialization}
To initialize the structure of understanding and initial confidence level on evidence randomly, the model generates a random float between 0 and 1 as confidence level for each evidence, so as for each non zero element of the structure of understanding that is chosen with polarization index $p = 0.5$. The sum of the structure of understanding is then normalized to 1. 

\begin{figure}[H]
    \centering
    \includegraphics[width=1\linewidth]{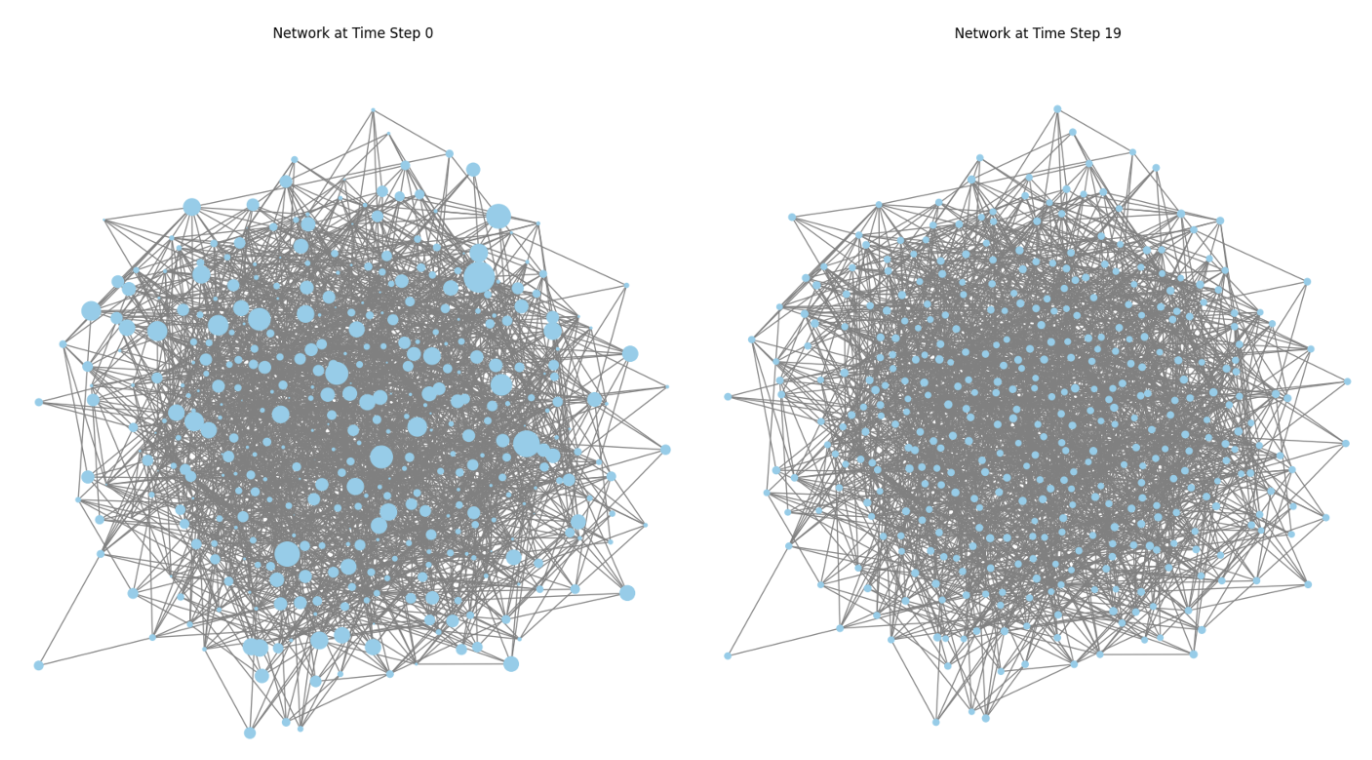}
    \caption{Initial and final belief network visualization for randomly generated structure of understanding and initial confidence level.}
    \label{fig:enter-label}
\end{figure}

The model then generates the belief distribution, network visualization and social pressure at each time step. In the network visualization, size of node represents level of agreement the agent has with the statement. A summary of results is listed below.

Observing the initial and final network visualization of belief distribution we can see the belief of the population has reached a consensus after 40 steps. 

By observing the three distributions in figure 3, even though people have very different structure of understanding, due to a well diffused information on evidence and under the influence of social norm, final belief distribution is converged into a range between $0.492$ and $0.508$. Social pressure of the population is small overall with a decreasing trend and maximum at only 0.01.

Belief distribution converges to an equilibrium state among the community within 5 time steps as we can see in figure 2. Social presure all start at 0, peaks at the begining then diminised after a while. Notice that the magnitude of social pressure increases with increasing change in the certainty of belief. This phenomenon is important in the next two experiments. 

After generating belief distribution and network visualizations, standard deviation of belief distribution versus connectivity index and population size are plotted respectively in figure 13. 

\begin{figure*}
\begin{minipage}{0.45\textwidth}
        \centering
        \includegraphics[width=\textwidth]{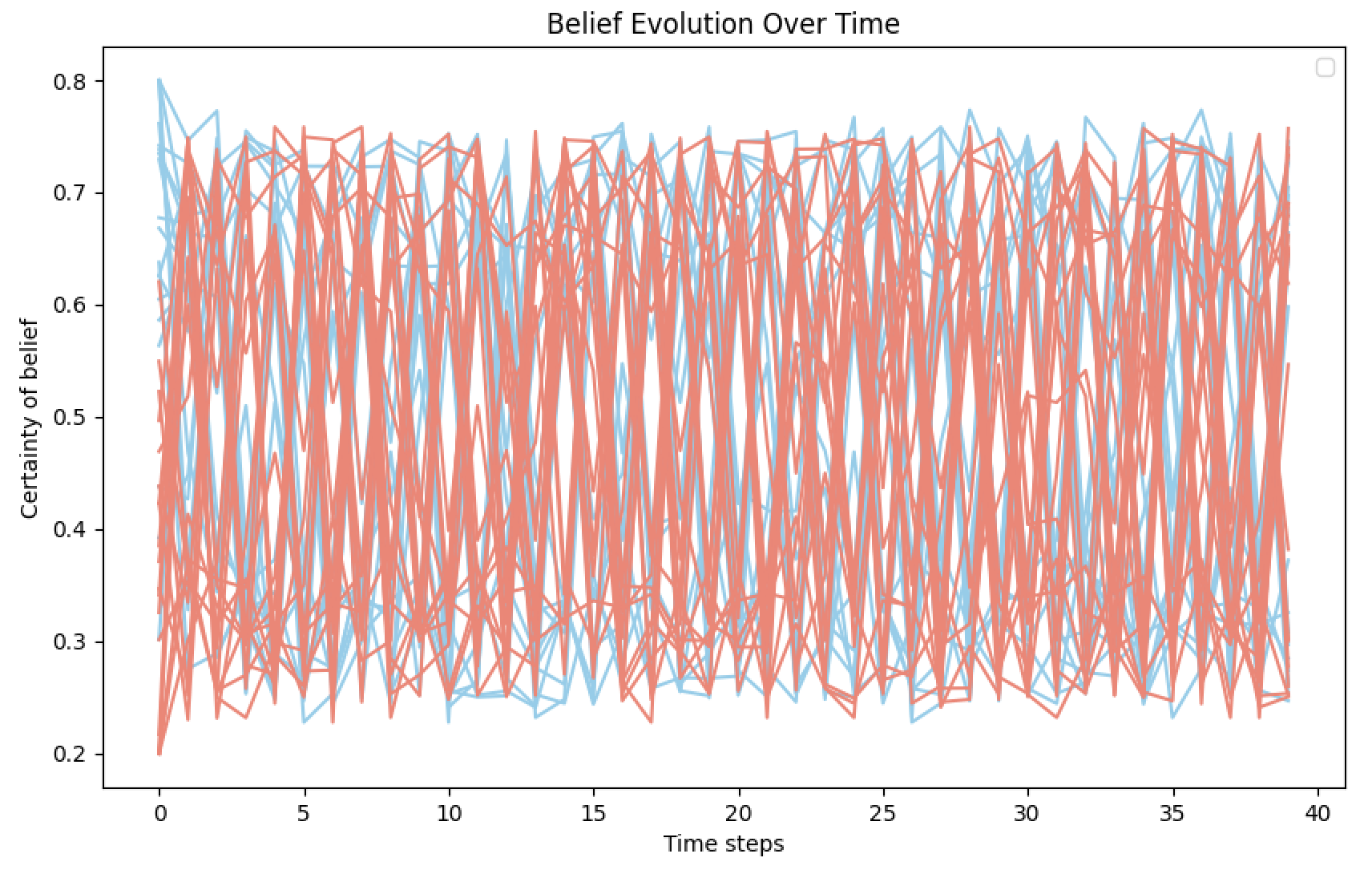} 
    \end{minipage}
    \begin{minipage}{0.45\textwidth}
        \centering
        \includegraphics[width=\textwidth]{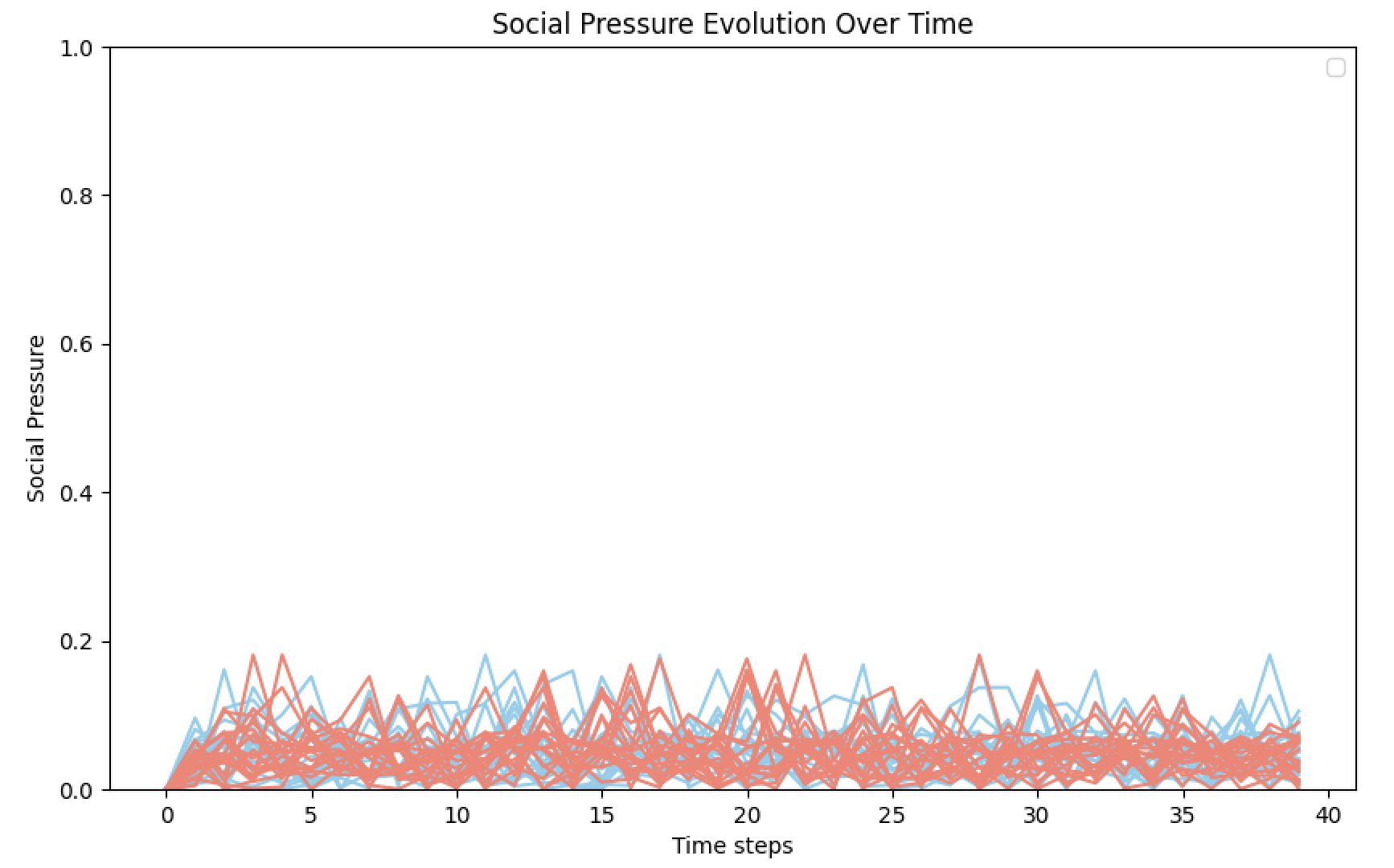} 
    \end{minipage}
    \caption{Evolution of belief (left) and social pressure (right) of the first and last 20 agents from each groups respectively over time for polarized structure of understanding and initial confidence level inside a network with two communities.}
    \label{fig:enter-label}
\end{figure*}
\begin{figure*}
    \centering
    \begin{minipage}{0.3\textwidth}
        \centering
        \includegraphics[width=\textwidth]{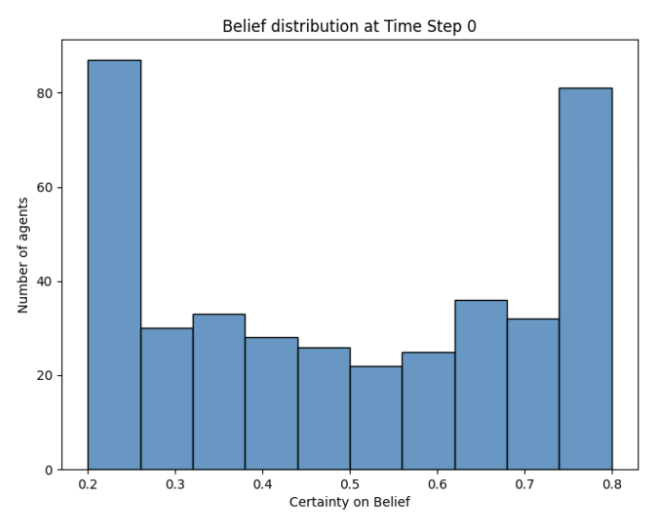} 
    \end{minipage}
    \begin{minipage}{0.3\textwidth}
        \centering
        \includegraphics[width=\textwidth]{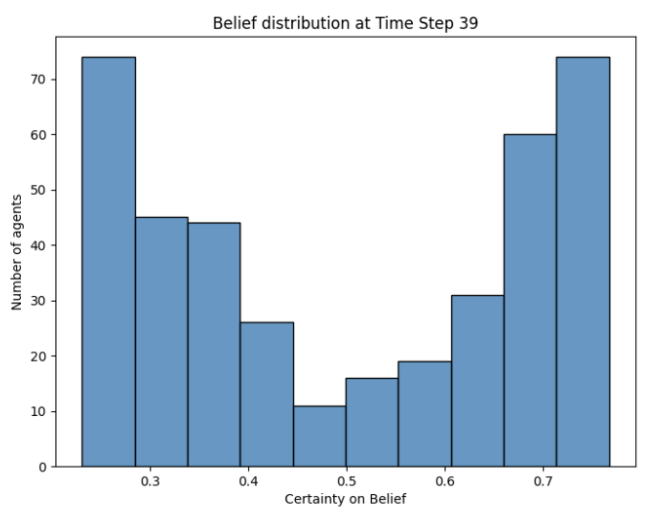} 
    \end{minipage}
    \begin{minipage}{0.3\textwidth}
        \centering
        \includegraphics[width=\textwidth]{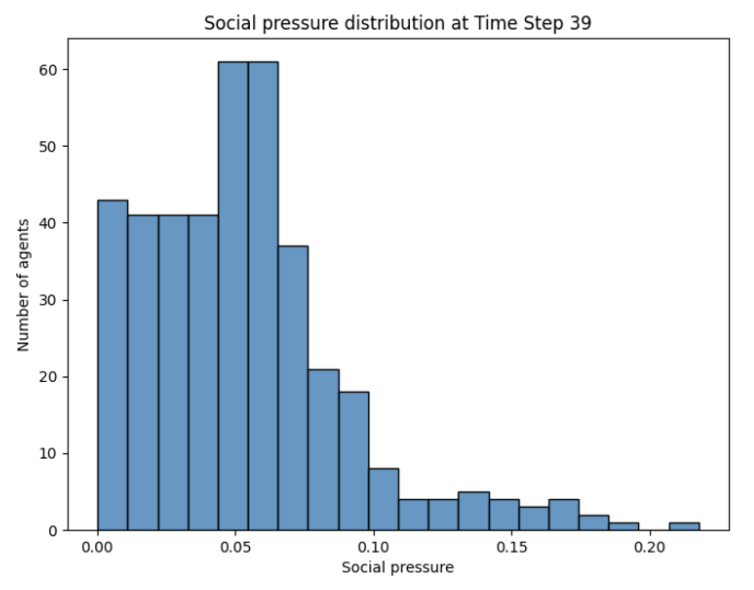} 
    \end{minipage}
    \label{fig:enter-label}
    \caption{Initial and final belief distribution and social pressure distribution.}
\end{figure*}

\subsection{Polarized Initialization}
\subsubsection{With Community Structure}
\begin{figure*}
    \centering
    \begin{minipage}{0.3\textwidth}
        \centering
        \includegraphics[width=\textwidth]{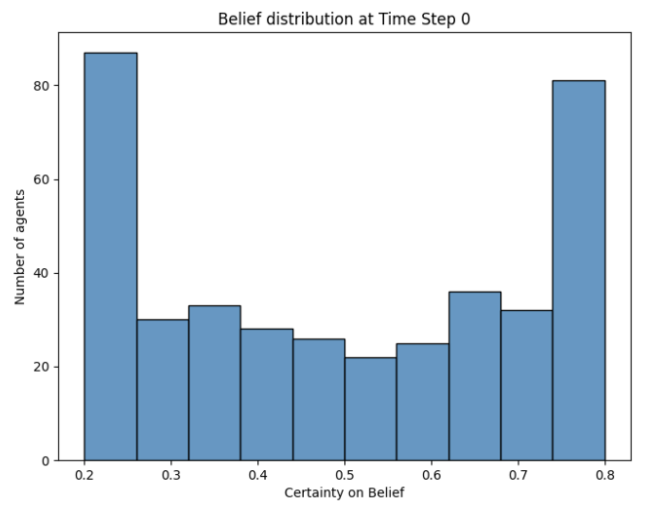} 
    \end{minipage}
    \begin{minipage}{0.3\textwidth}
        \centering
        \includegraphics[width=\textwidth]{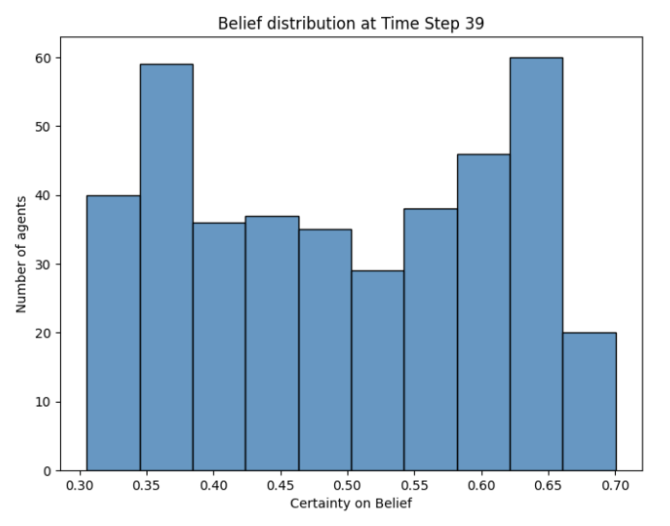} 
    \end{minipage}
    \begin{minipage}{0.3\textwidth}
        \centering
        \includegraphics[width=\textwidth]{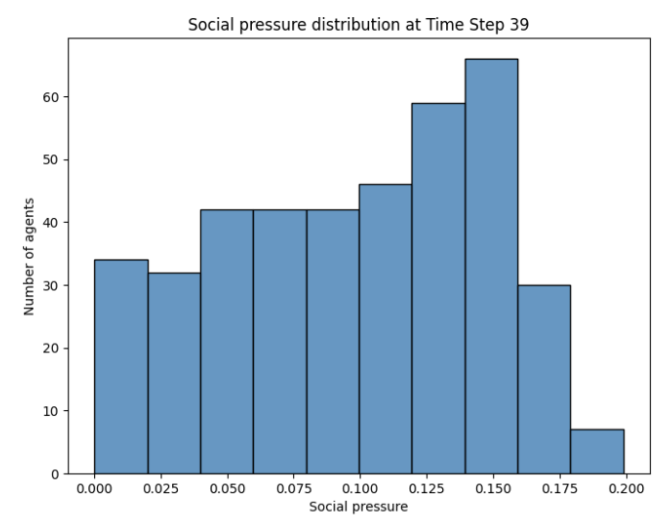} 
    \end{minipage}
    \label{fig:enter-label}
    \caption{Initial and final belief distribution and social pressure distribution for polarized structure of understanding and initial confidence level inside a giant component.}
\end{figure*}
\begin{figure*}
\begin{minipage}{0.45\textwidth}
        \centering
        \includegraphics[width=\textwidth]{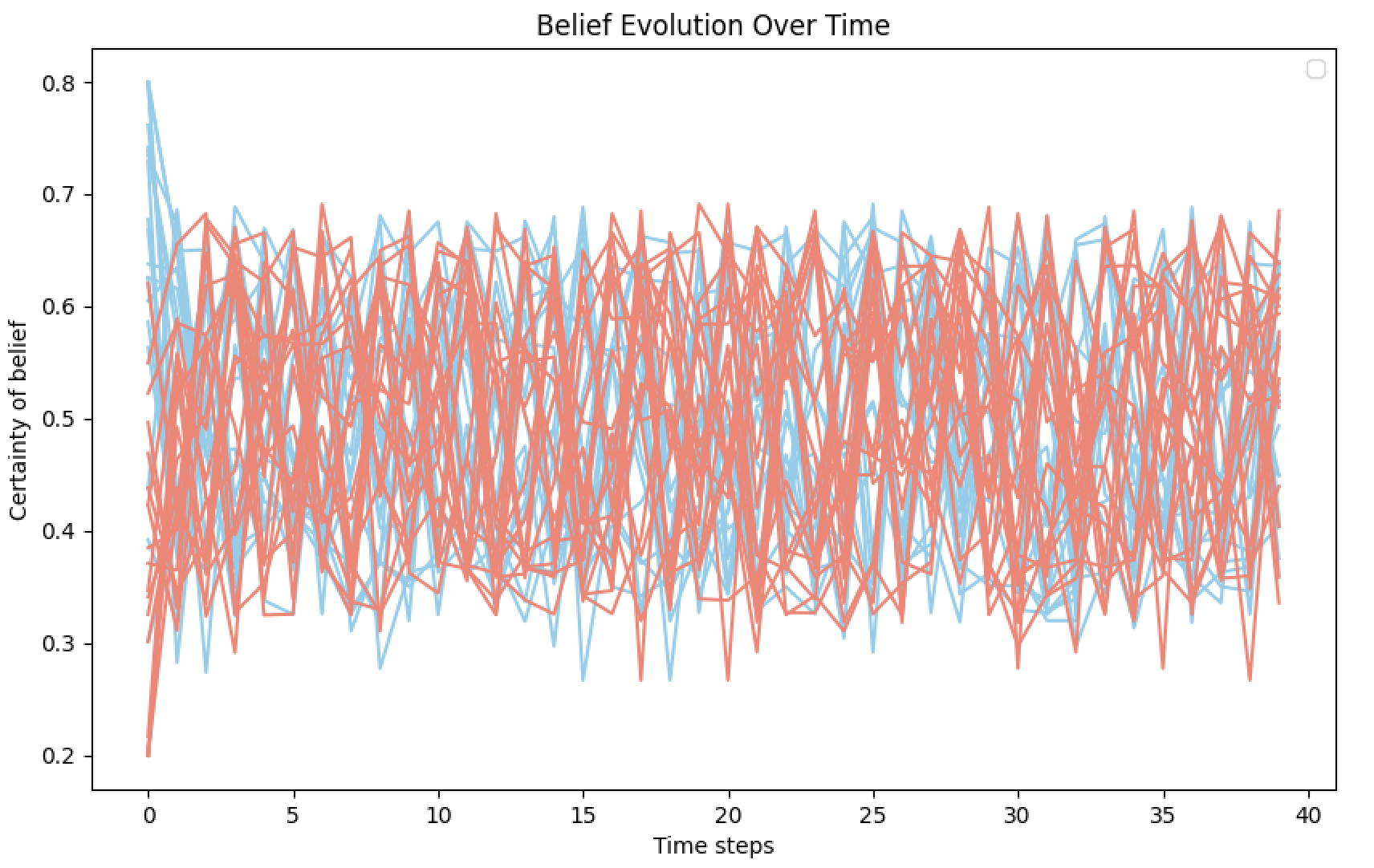} 
    \end{minipage}
    \begin{minipage}{0.45\textwidth}
        \centering
        \includegraphics[width=\textwidth]{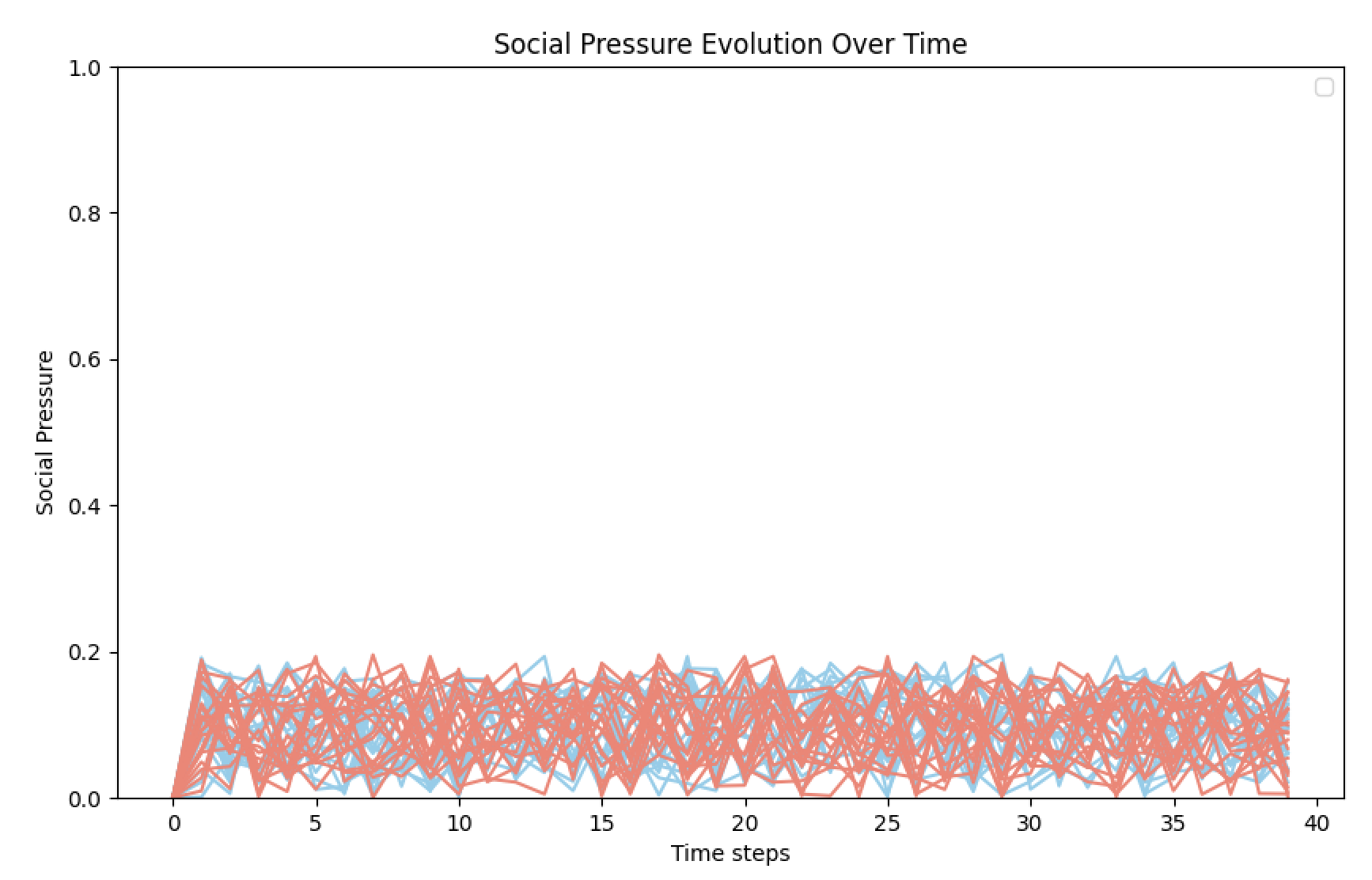} 
    \end{minipage}
    \caption{Evolution of belief (left) and social pressure (right) of the first and last 20 agents from each polarized groups respectively over time in a giant component.}
    \label{fig:enter-label}
\end{figure*}

To generate polarized structure of understanding and initial confidence level on evidence, we use the same method as in random initialization, but with polarization index $p=0.8$. Initial confidence level is generated with half of the population with confidence level $0.2$ on the first $5$ evidence, $0.8$ on the second $5$ evidence, and half of the population with confidence level $0.8$ on the first $5$ evidence, $0.2$ on the second $5$ evidence. A network with two clear equal sized community structure is constructed.

The model then generates the belief distribution, network visualization and social pressure at each time step. In the network visualization, size of node represents level of agreement the agent has with the statement, just like before, and red and blue colors are used to denote the two polarized groups respectively. We also show the social pressure network visualization with the same idea.

By observing the network visualization of belief distribution in figure 7, we can see that initially the blue community tends to belief in the statement more than the red community, where blue and red represent communities that lean toward positive and negative side of the evidence respectively. After the distribution converge, belief dynamics of each community starts to fluctuate while the overall belief distribution is the same. There is about equal amount of individuals believing in the statement as there is not in each community. This can also be reflected in the belief evolution and social pressure evolution graph in figure 5 and figure 6. After moderate convergence, belief of individuals start to oscilate out of phase with social pressure. This is consistent with our understanding of social pressure being the part of the drive force of belief change. 

\begin{figure}[H]
    \centering
    \includegraphics[width=1\linewidth]{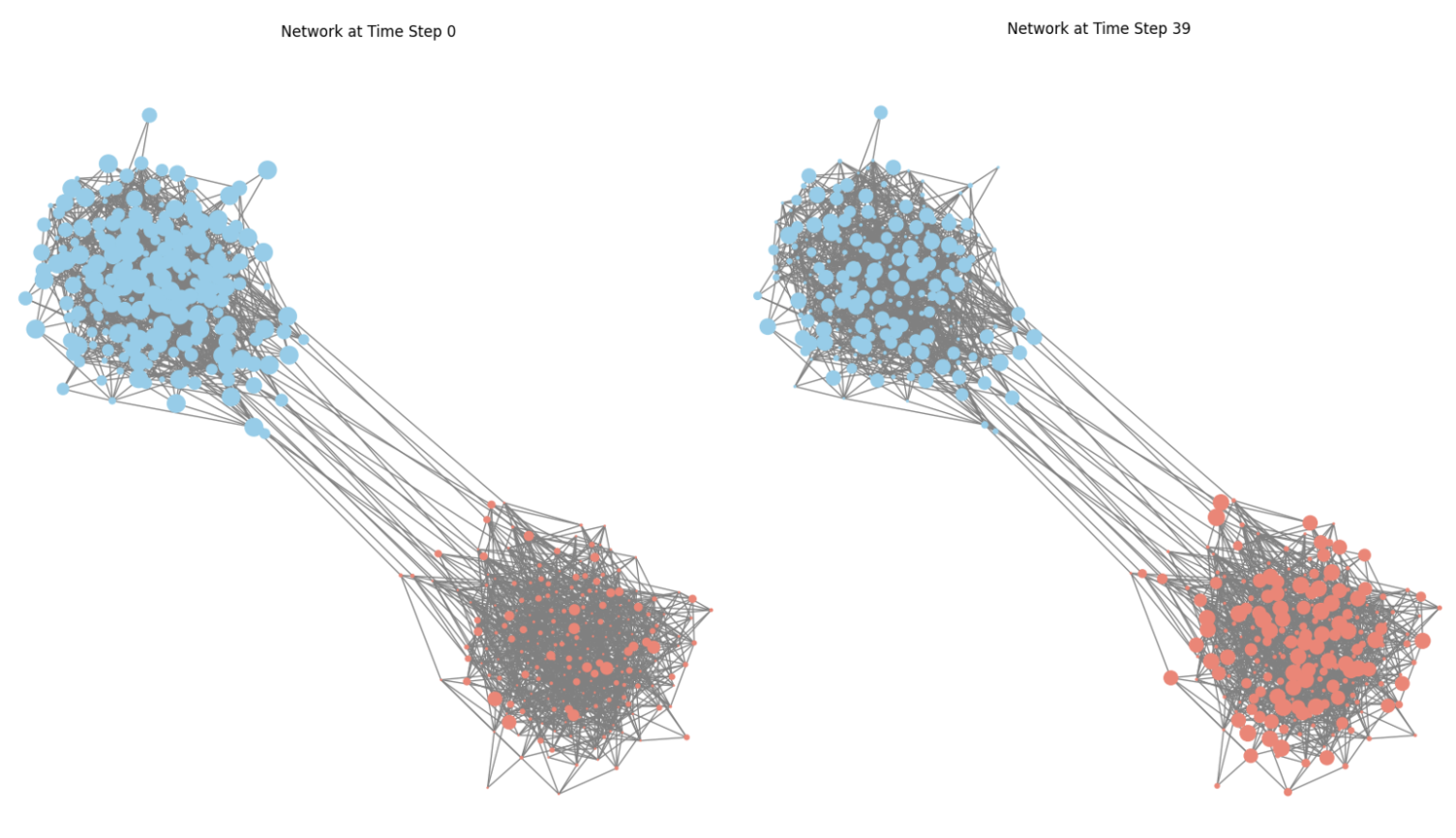}
    \caption{Initial and final belief network visualization for polarized structure of understanding and initial confidence level on a network with two communities.}
    \label{fig:enter-label}
\end{figure}

Figure 8. shows us the social pressure distribution in the network after belief distribution reaches equilibrium. The location of large social pressure seems to have no preference over connecter nodes and cluster nodes.

\begin{figure}[H]
    \centering
    \includegraphics[width=1\linewidth]{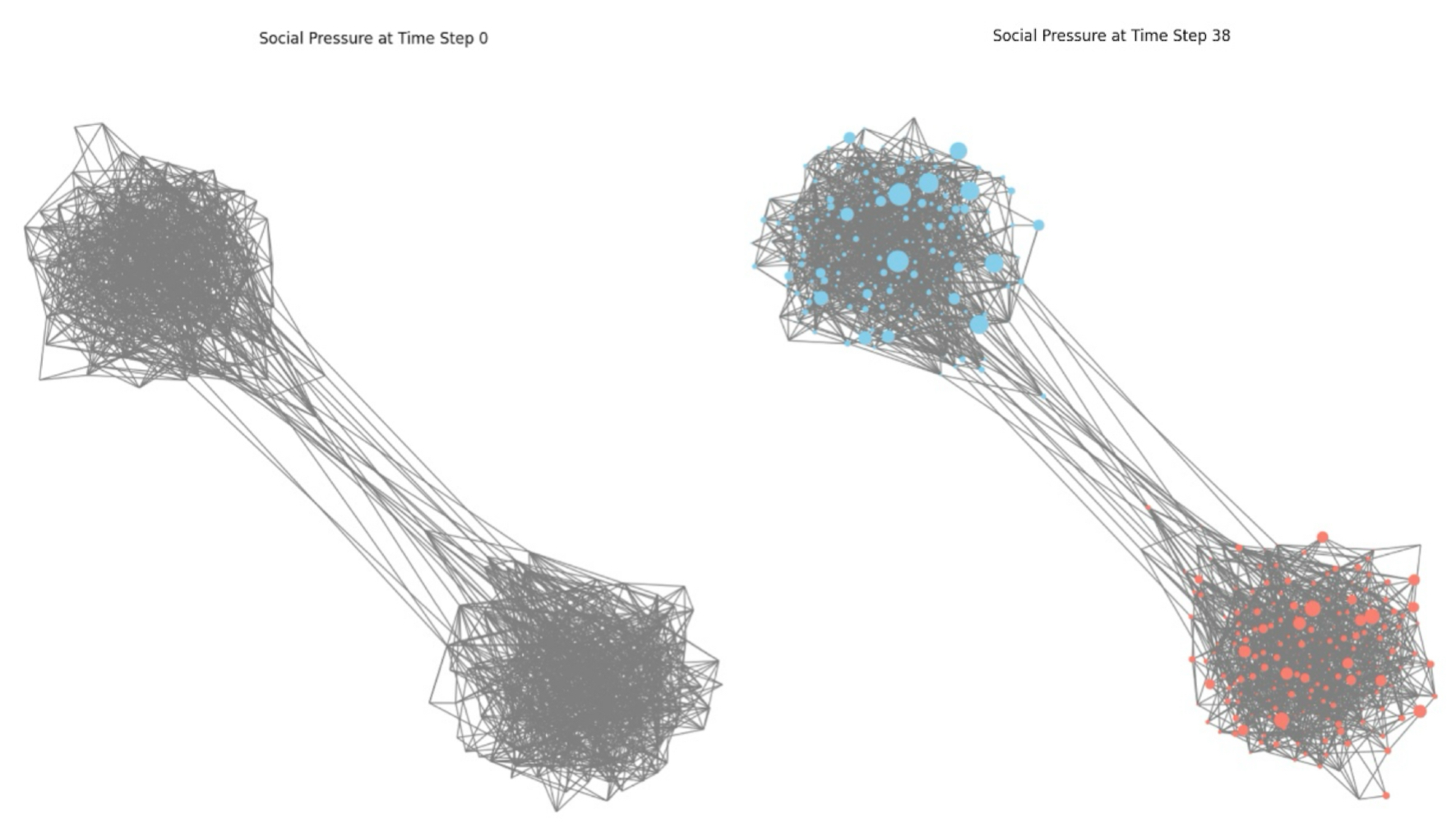}
    \caption{Initial and final social pressure network visualization.}
    \label{fig:enter-label}
\end{figure}

Initially, polarization exist in separate community, and over time it moves into each community. Previous research has shown that intraparty polarization does exist in american politics \cite{Groenendyk2020}. Evidence propagation could be a cause. In figure 6, Social pressure distribution shows a decreasing trend with more fraction of the population on the lower end of the social pressure spectrum than the randomly generated case and also a few at higher. 

\subsubsection{With Giant Component}
The polarized structure of understanding and initial confidence level on evidence is constructed in the same way as before, but this time implemented on a network with a giant component.

In the network visualization, figure 11, we still denote the two communities with red and blue color respectively, but this time randomly distributed inside a network with a giant component. According to figure 9 and 10, convergent effect still exists, weaker than the random initialization but stronger than a polarized network with community structure.

Social pressure distribution although have similar maximum as the distribution with community structure at $0.2$, we see an increasing trend until a few at high pressure instead of a deceasing trend. Agents inside a giant component has a more uniform belief distribution, but it has a much higher social pressure than in a network with community structure. This might be the explaination why people tend to form community instead of a giant component as a way to reduce tension. 

Further analysis on polarization is shown in figure 14.

\begin{figure}[H]
    \centering
    \includegraphics[width=1\linewidth]{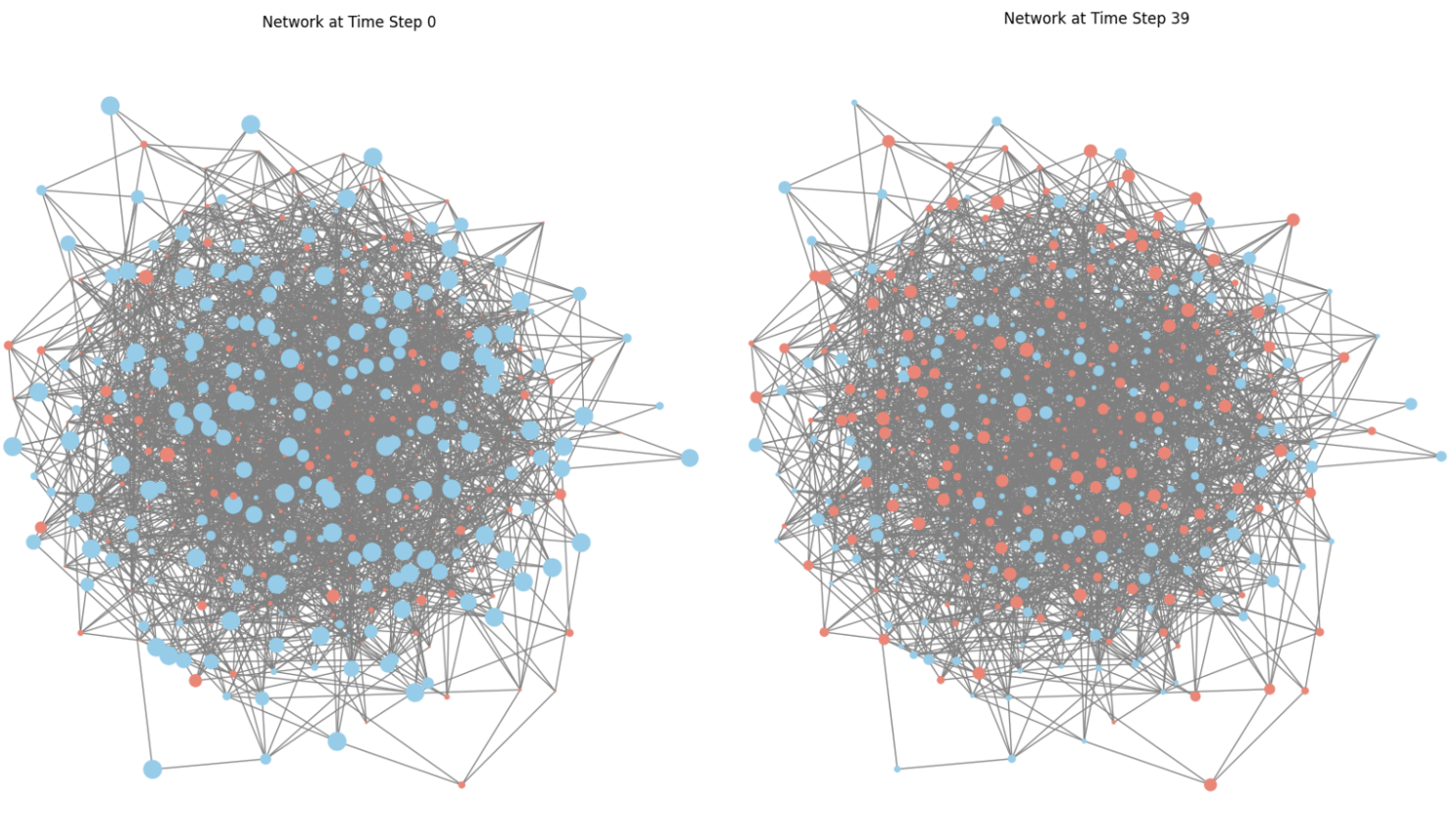}
    \caption{Initial and final belief network visualization for polarized structure of understanding and initial confidence level the same as before but inside a giant component.}
    \label{fig:enter-label}
\end{figure}
\begin{figure}[H]
    \centering
    \includegraphics[width=1\linewidth]{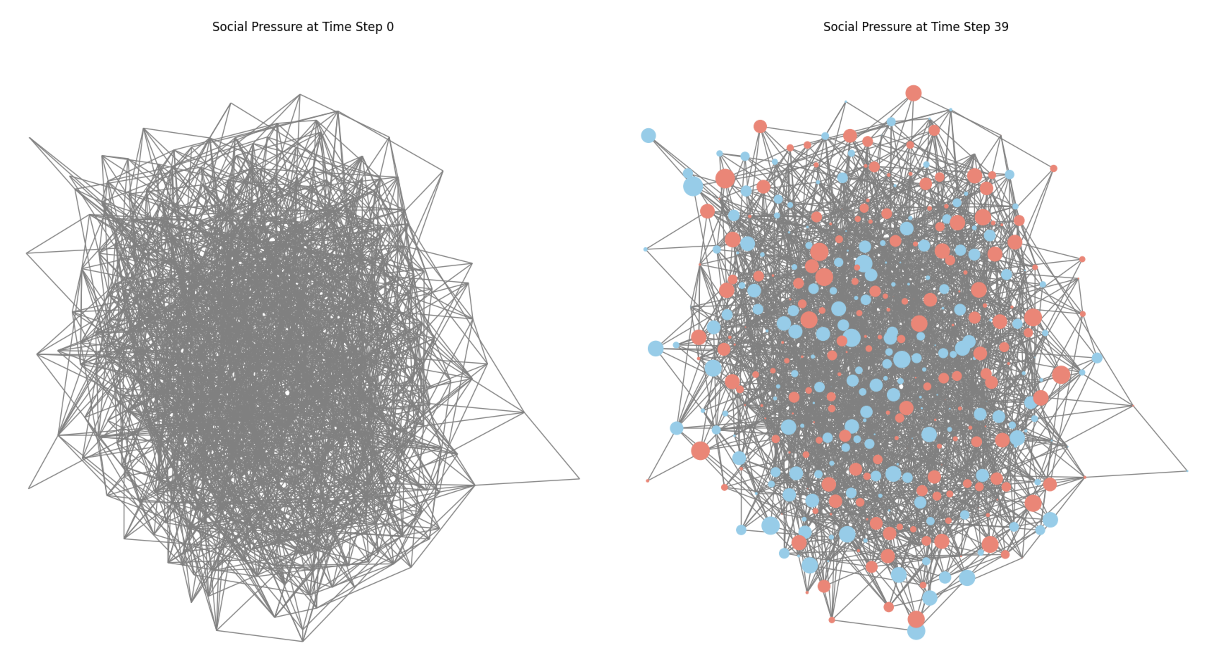}
    \caption{Initial and final socail presure network visualization.}
    \label{fig:enter-label}
\end{figure}

\section{Further Analysis of Model Parameters}
\begin{figure*}
    \centering
    \begin{minipage}{0.49\textwidth}
        \centering
        \includegraphics[width=\textwidth]{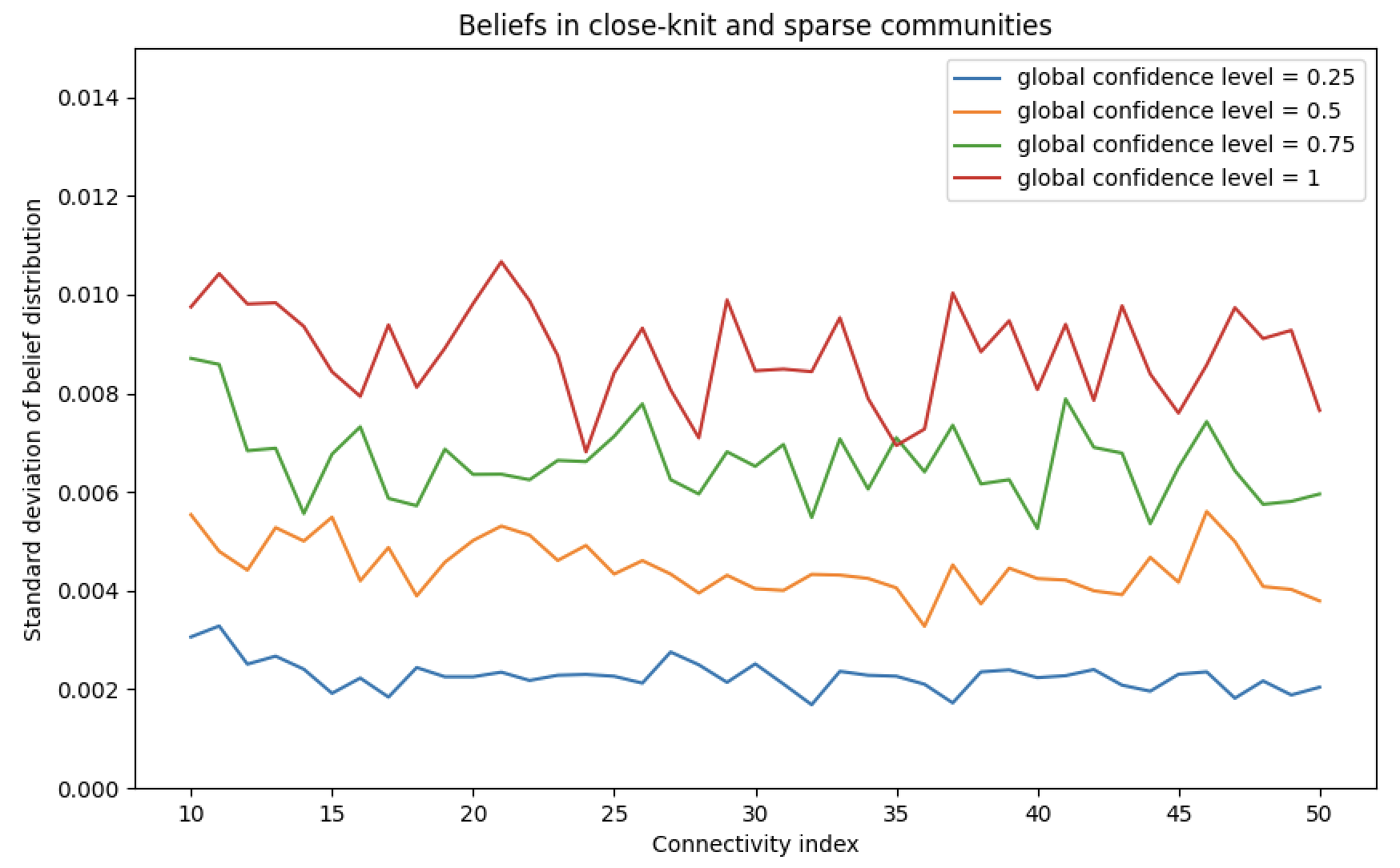} 
    \end{minipage}
    \begin{minipage}{0.49\textwidth}
        \centering
        \includegraphics[width=\textwidth]{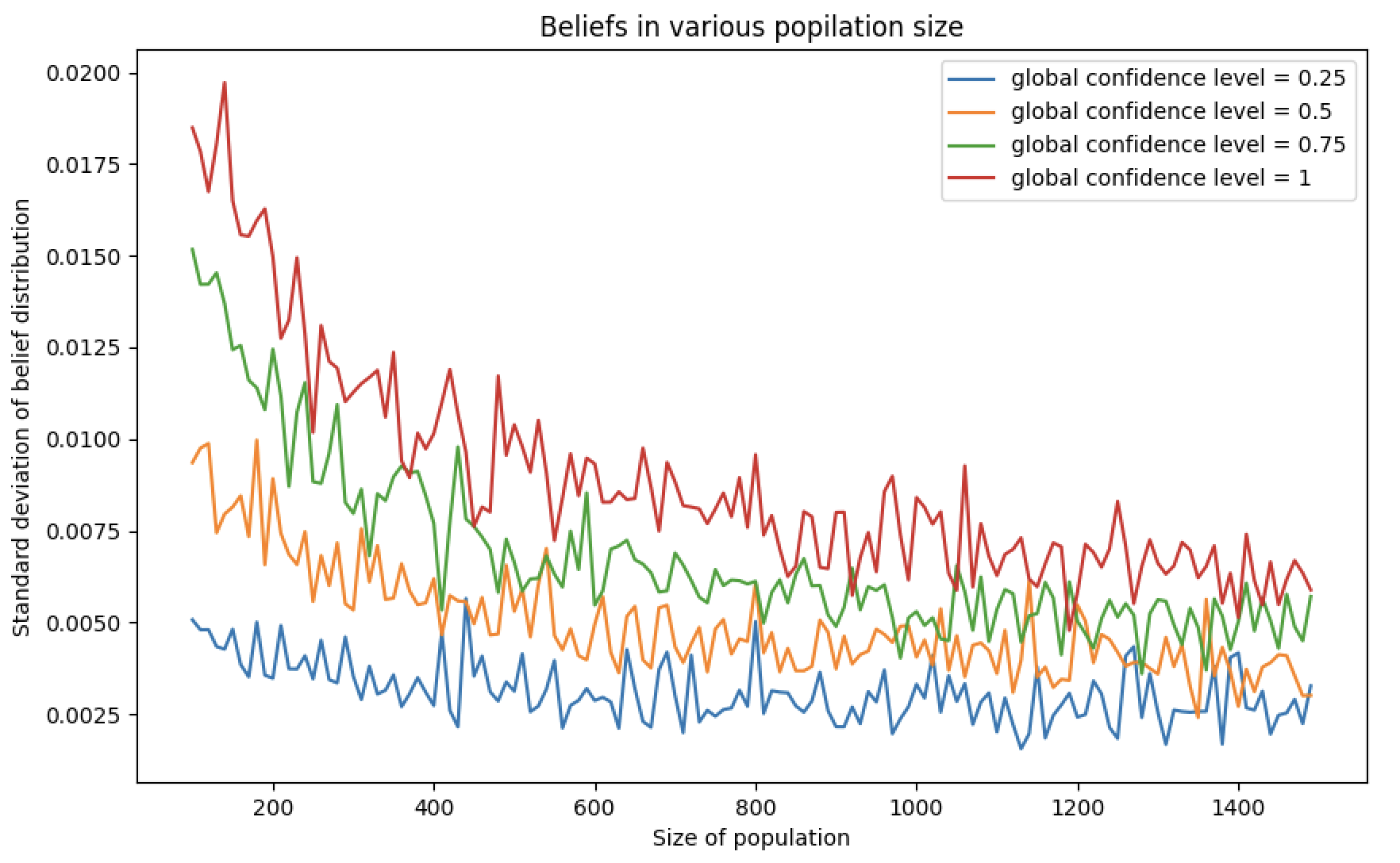} 
    \end{minipage}
    \caption{A plot of standard deviation of belief distribution versus connectivity index $k$ for a population of 400 people (left figure) and a plot of standard deviation of belief distribution versus population size with connectivity $k=10$ (right figure) at various self-confidence level under random initialization of structure of understanding is shown above.}
\end{figure*}
\begin{figure*}
    \centering
    \begin{minipage}{0.49\textwidth}
        \centering
        \includegraphics[width=\textwidth]{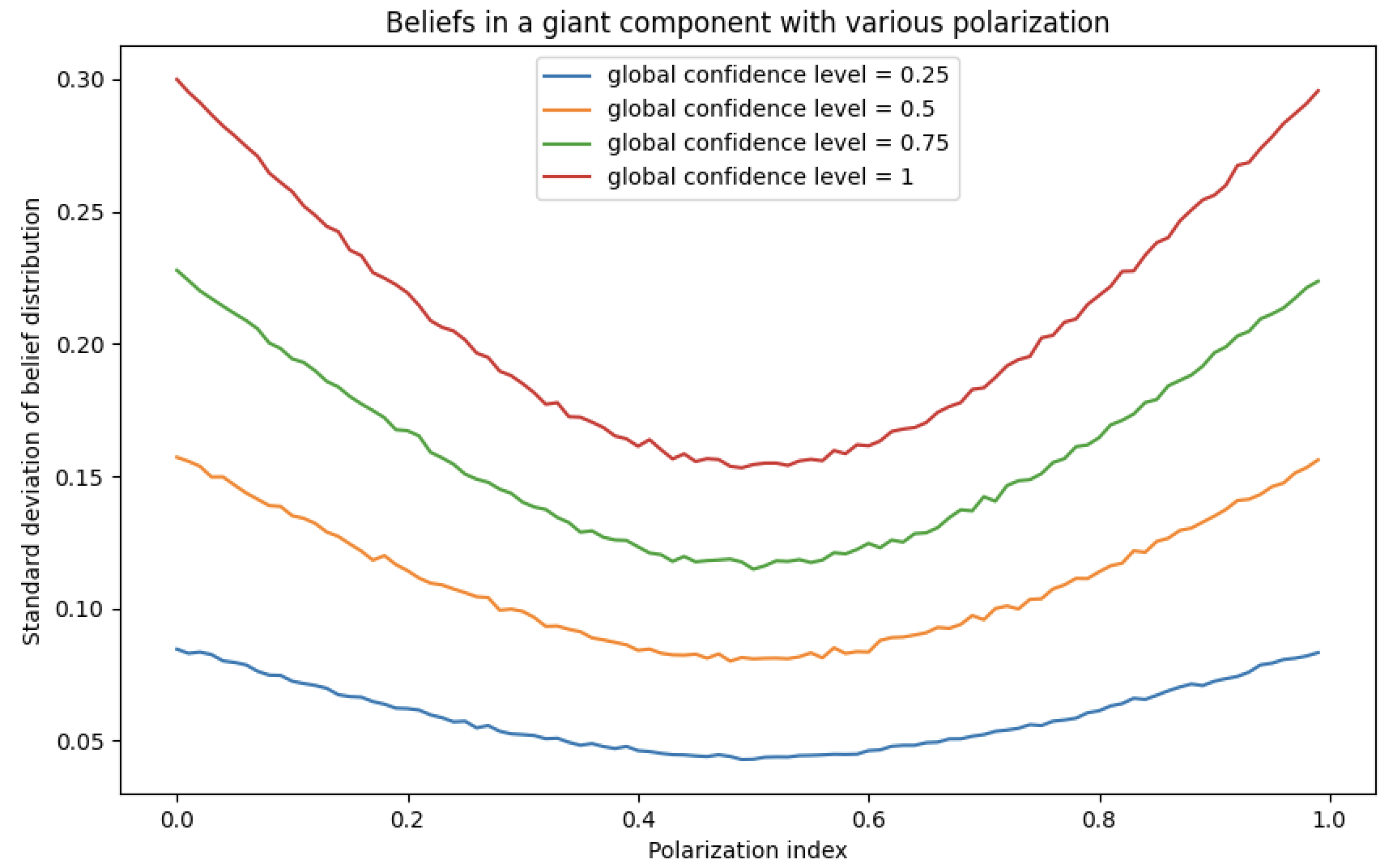} 
    \end{minipage}
    \begin{minipage}{0.49\textwidth}
        \centering
        \includegraphics[width=\textwidth]{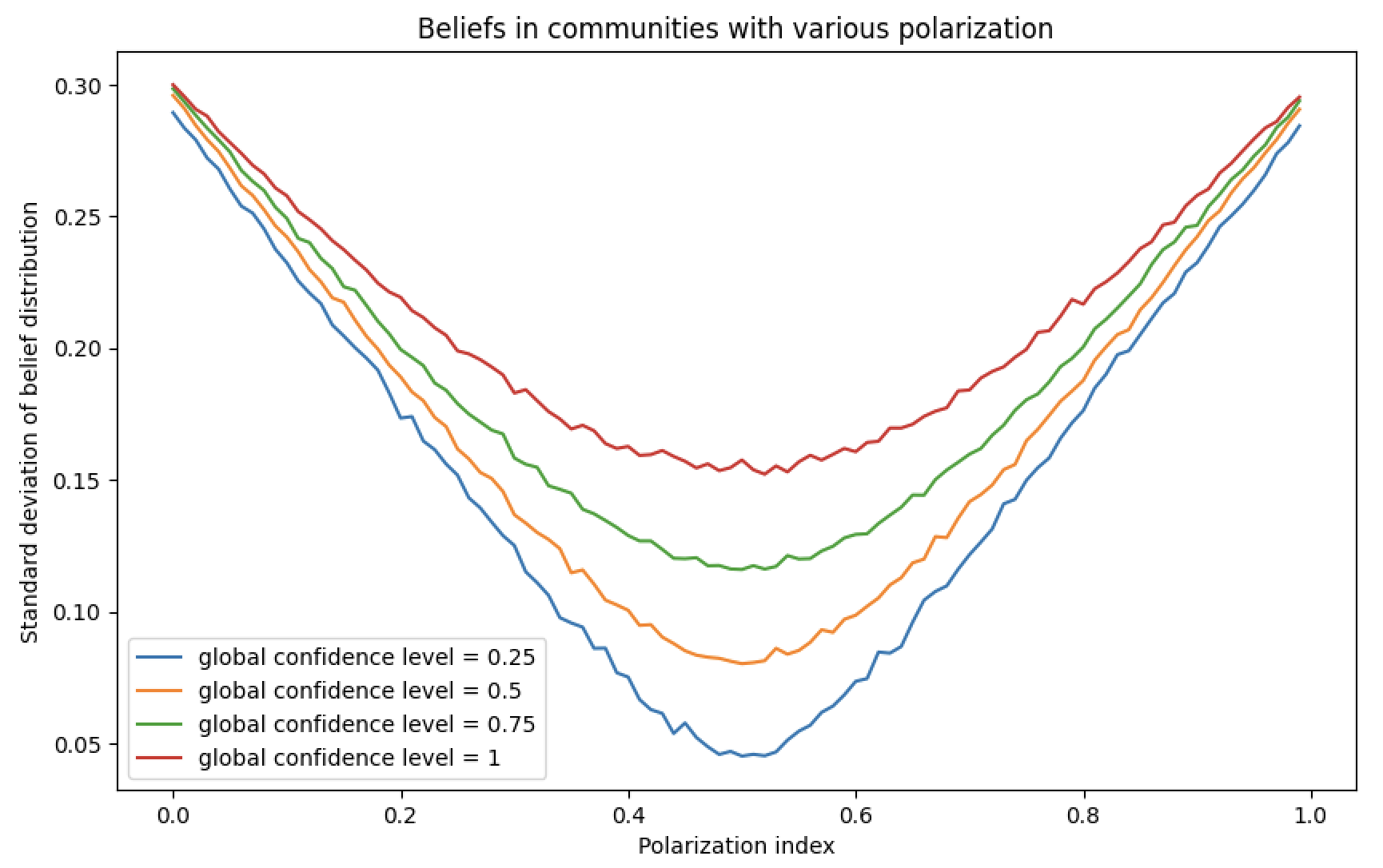} 
    \end{minipage}
    \caption{Plots of standard deviation of belief distribution versus polarization index $p$ for a polarized population with a giant component (left figure) and with a community structure (right figure) with a population size of 400 and connectivity index $k = 10$ at various self-confidence level is shown above.} 
\end{figure*}
\begin{figure*}
    \centering
    \begin{minipage}{0.49\textwidth}
        \centering
        \includegraphics[width=\textwidth]{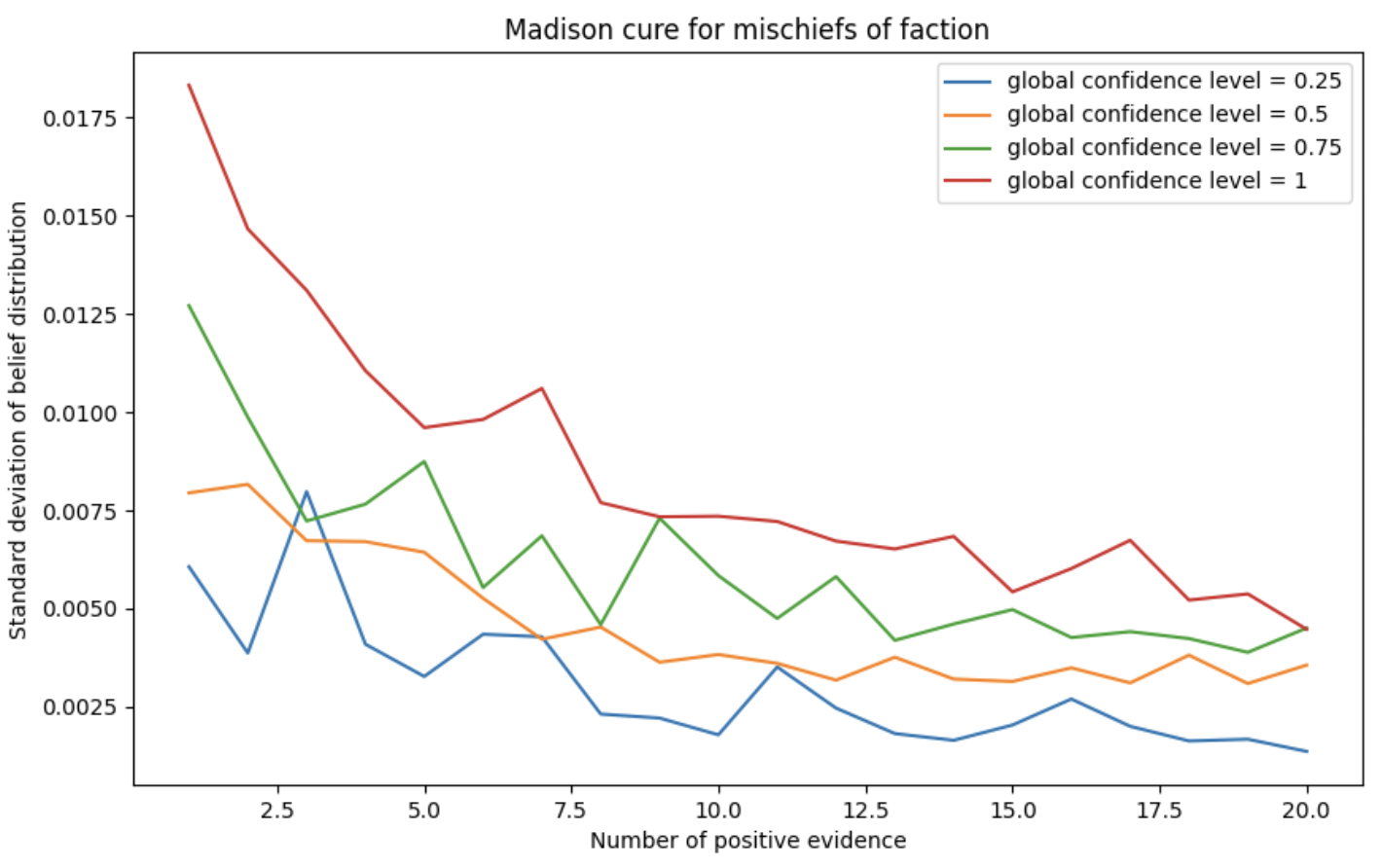} 
    \end{minipage}
    \caption{Plots of standard deviation of belief distribution versus number of positive evidence $m$ for a population of 400 people, connectivity index $k = 10$ at various self-confidence level under random initialization of structure of understanding is shown above.} 
\end{figure*}
We also investigated how connectivity idex, polulation size and polarization index in different network change the variation in equilibrium belief distribution.

According to figure 13, connectivity is independent of standard deviation in the belief distribution over connectivity index bigger than 10 (since for $k<10$ the giant component can't cover enough fraction of the population, leading to isolated agents that increases variance significantly), and the standard deviation in the belief distribution decreases with increasing population size. This is a surprising phenomenon, since local structure of the network is the same throughout different population size. Standard deviation of belief distribution increases with confidence level as expected.

To see how different levels of polarization in different community structures influence belief distribution, we performed the experiment with result shown in figure 14. The higher the confidence level, the larger $p$ or $1-p$ is, the larger the standard deviation in the belief distribution. When self-confidence level is 0, consensus is reached, since the model reduced to DeGrootian, standard deviation is simply zero. When confidence level is $1$, network structure does not influence individual's decision based on polarization, as the shape of both curves look the same. Both plots seem to share the same value across different self-confidence level at polarization index $p=0.5$, which indicates no poarlization. When polarization index goes to the extremes, network with community structure is less sensitive to change in confidence level than network with a giant component. 

By introducing the structure of understanding, we can also investigate the effectiveness of Madison's cure for the mischiefs of faction, mentioned in Federalist 10, by letting the statment to be level of support to the government, and evidence to be the issues citizens can care about to support the government. The more common faith people have in their government, the less conflicts can happen between factions. In figure 15, an increase of the diversity of evidence one can consider for his or her belief causes consensus to reach more effectively. Previous model has demonstrated this conclusion \cite{Kawakatsu2021}.

\section{Discussion}
By introducing the concept of structure of understanding and evidence into the Friedkin-Johnsen model, we are able to probe the underlying cognitive structure of belief propagation further. In the numerical experiment, two network structures, one with a giant component and one with two communities are explored. In a polarized population, standard deviation of belief distribution is smaller for a network with a giant component than a network with two community structure, but the former network has a larger social pressure than the later network. In a network of a giant component with randomly generated structure of understanding and confidence level on evidence, the larger the population size, the smaller the variation in belief distribution. Connectivity index does not influence the variation in belief distribution. 

Structures of understanding itself can also be subject of influence. One possible way to update structure of understanding is through a similar process from Axelrod’s Culture Model, where after certain time step, each person will exchange a small portion of their structure of understanding with their neighbor’s. Another way, which might be the most important way to change people’s structure of understanding is through pressure from the environment that causes the person to undergo a reinforcement learning process. This pressure could be social pressure, or one can potentially develope a new functioning environment that can interact with the agent’s structure of understanding.

The weight on the social network can also be subject to change due to cognitive dissonance. A model that allows social network weights to change with respect to cognitive dissonance has been developed\cite{Galesic2021}, which can be potentially combined with this model.

Currently we are considering independent evidences, which means receiving one evidence has no inflence on the certainty of others. One future improvement on the model could be to model each individual as a multilayered neural network instead of just one layer, with each evidence being a statement of its own. The deeper the neural network, the higher the resolution of the belief. Propagation of multiple statements instead of one can also be implemented.

\begin{acknowledgments}
I wish to thank the University of Michigan Department of Mathematics to provide the opportunity for me to conduct this research in cmplxsys 270. 

I would also like to thank the developers of mesa package for agent based simulation in python.\cite{python-mesa-2020}
\end{acknowledgments}
\vspace{1cm}
\appendix

\section{Github repository}
You can access the code of this project by cloning the following repository. 

https://github.com/jeffyujianfu/BPN.git

\bibliography{reference}

\end{document}